\documentclass[preprint,11pt]{elsarticle}

\usepackage{ucs}
\usepackage[left=2.5cm,right=2.5cm,top=2.5cm,bottom=2.5cm]{geometry}
\usepackage[T1]{fontenc}
\usepackage[utf8x]{inputenc}
\usepackage{amsmath}
\usepackage{multirow}
\usepackage[]{graphicx}
\graphicspath{{./rysunki/}{./rysunki.src/}} 
\usepackage{pslatex}

\usepackage[dvips]{hyperref}
\usepackage{breakurl}

\usepackage{array}

\begin{document}
\sloppy
\newcommand{\LinkThroughput}{R_{link}}
\newcommand{\MaxAckLatency}{t_{ack~max}}
\newcommand{\BufferCapacitance}{M_{buf}}
\newcommand{\TransmSpeed}{R_{transm}}
\newcommand{\NOfPktBuffers}{N_{pkts}}
\newcommand{\LogNOfPktBuffers}{N_{p~bits}}
\newcommand{\RatioResent}{R_{rsnt}}
\newcommand{\NOfSentPackets}{C_{pkt~sent}}
\newcommand{\NOfResentPackets}{C_{pkt~rsnt}}
\newcommand{\NOfCountedPackets}{N_{pkt~update}}
\newcommand{\NCAHighThreshold}{T_{high}}
\newcommand{\NCALowThreshold}{T_{low}}
\newcommand{\NCADecreaseDelay}{\alpha_{decr}}
\newcommand{\NCAIncreaseDelay}{\alpha_{incr}}

\newcommand{\WZcode}[1]{{\em #1}}
\newcommand{\WZhex}[1]{{\em #1}}

\begin{frontmatter}
\title{Efficient transmission of measurement data from FPGA to embedded system via Ethernet link}
\author{Wojciech M. Zabolotny\corref{a1}}
\cortext[a1]{Tel. +48 22 234 7717; fax.: +48 22 825 2300}
\ead{wzab@ise.pw.edu.pl}
\address{Institute of Electronic Systems, Warsaw University of Technology,
 ul. Nowowiejska 15/19, 00-665 Warszawa, Poland}

\begin{abstract}
This paper presents a system consisting of the FPGA IP core, the simple
network protocol and the Linux device driver,
capable of efficient and reliable data transmission
from a low resources FPGA chip to the Linux-based embedded computer system, via a private
Ethernet network (consisting of a single segment or
a few segments connected via an Ethernet switch).
The embedded system may optionally process the acquired data, and distribute them
further, using standard network protocols.

Proposed design targets cost-efficient multichannel data acquisition systems,
in which multiple FPGA based front end boards (FEB) should 
transmit the stream of acquired data to the computer network,
responsible for their final processing and archiving.

The presented solution allows to minimize the cost of data concentration
due to use of inexpensive Ethernet network infrastructure.

The work is mainly focused on minimization of resources consumption in the FPGA,
and minimization of acknowledge latency in 
the Linux based system - which allows to achieve high throughput 
in spite of use of inexpensive FPGA chips with small internal memory.
\end{abstract}
\end{frontmatter}
\section{Introduction}
\begin{figure}[t]
   \begin{center}
   \begin{tabular}{c}
   \includegraphics[width=0.8\linewidth]{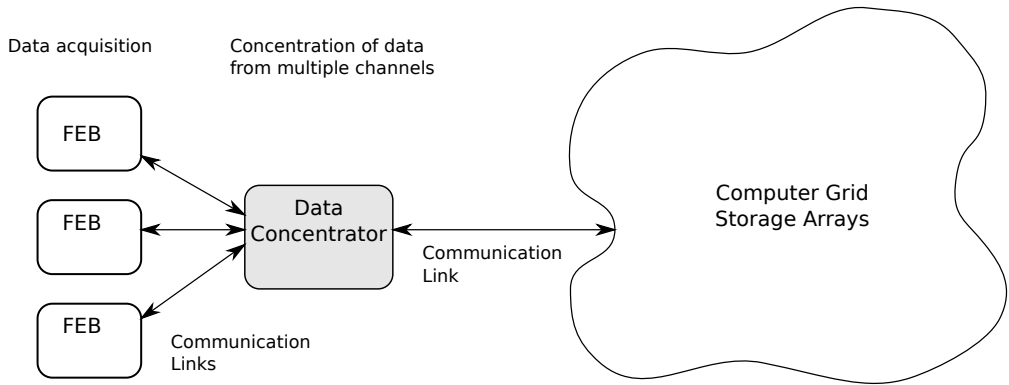}
   \end{tabular}
   \end{center}
   \caption[example]
   { \label{figure:gen-daq-1}
     Data acquisition system with multiple FEBs connected via data concentrator 
     to the data processing grid.
   }
\end{figure}

Contemporary measurement systems often are spatially distributed and use multiple
input channels to acquire data. 
The designers of the data acquisition (DAQ) part of such system have to solve
the problem how to transmit data from multiple front end boards (FEB), 
receiving signals from the sensors and converting them into the digital form,
to the data processing center (often a computer grid or storage array),
which will finally process and archive those data (see Figure~\ref{figure:gen-daq-1}).
This process involves concentration of data, often associated 
with preprocessing or aggregation of  concentrated data.
In systems with high numbers of input channels, the cost of the data concentration
subsystem may significantly affect the cost of the whole DAQ, and therefore 
development of cost-efficient solutions is desirable.

The digital part of FEB boards is often based on the~Field Programmable Gate Arrays
(FPGA) chips. Such solution offers many advantages:
\begin{itemize}
 \item high flexibility, and possibility to adjust or correct data 
  acquisition algorithms without hardware modifications
 \item good performance of simultaneous processing of parallel data streams
 \item easy interfacing to different digital interfaces, 
       used to connect analog to digital converters (ADC)
       or sensors with digital outputs.
 \item deterministic timing and data sampling with minimized jitter
\end{itemize}
Because the FEBs are usually the most numerous components of the DAQ, 
it is desirable to minimize the cost of FPGA chips used in a FEB.

\begin{figure}[t]
   \begin{center}
   \begin{tabular}{c}
   \includegraphics[width=0.8\linewidth]{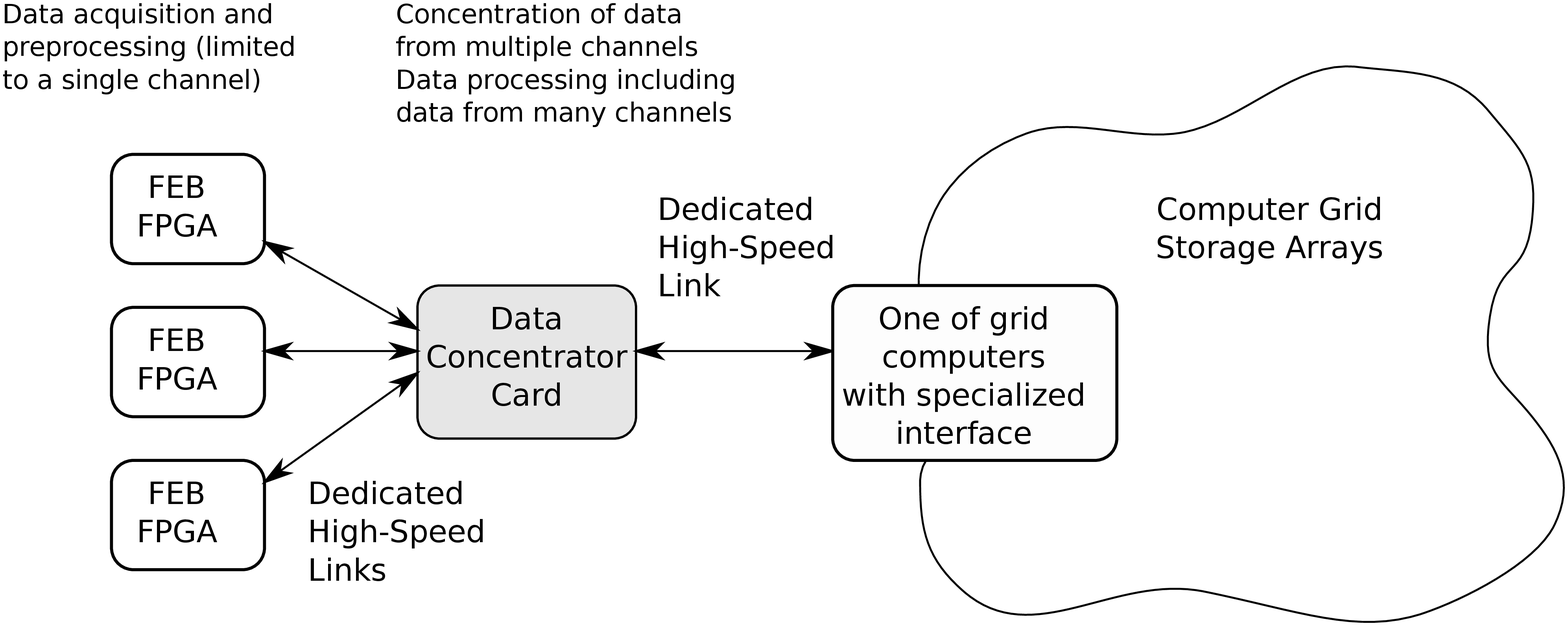}
   \end{tabular}
   \end{center}
   \caption[example]
   { \label{figure:dcc-daq-1}
     Data acquisition system with multiple FEBs connected via dedicated high-speed links to
     the specialized DCC board, and further, via another high-speed link to
     one of computers in the data processing grid.
   }
\end{figure}

Sometimes the process of data concentration may involve computation intensive
data processing using highly parallelized algorithms, and in this case the FPGA
based Data Concentration Card (DCC) may be needed (see Figure~\ref{figure:dcc-daq-1}). 
In this situation the high speed serial links (currently available in more advanced
FPGA chips\cite{aria-ii-transceivers, xilinx-transceivers}) may be a good solution
to provide transmission of data from the FEBs to the 
dedicated DCC. An example of such approach may be the DCC board used 
in the Resistive Plate Chamber Pattern Comparator Muon Trigger 
in the CMS Experiment\cite{Almeida:692739, art-wz-iop2007}. The problem of such approach
however is relatively high cost of development and manufacturing of such specialized
DCC board. The cost may be even further increased by high cost of specialized components 
needed to provide connection between the FEBs and DCC.

In some cases (e.g. when only data concentration is needed, or when the preprocessing 
of concentrated data involves mainly sequential algorithms), there is no advantage to use the
DCC card, and it may be replaced with an embedded computer system with performance
suitable to handle the expected data stream. Because such embedded systems may be equipped with
multiple network interfaces, and because it is possible to connect the FPGA in FEB to the network 
physical layer interface chip (PHY), the data concentration may be performed using a standard
network connection.
If the throughput of the network link is sufficient, we can use the network switch to connect
multiple FEBs to a single network interface.

Such approach, based on widely available, ready to use hardware solutions (network
cables, network switches, network capable embedded systems) may significantly decrease
the total cost of the system, and shorten the development time.

\begin{figure}[t]
   \begin{center}
   \begin{tabular}{c}
   \includegraphics[width=0.9\linewidth]{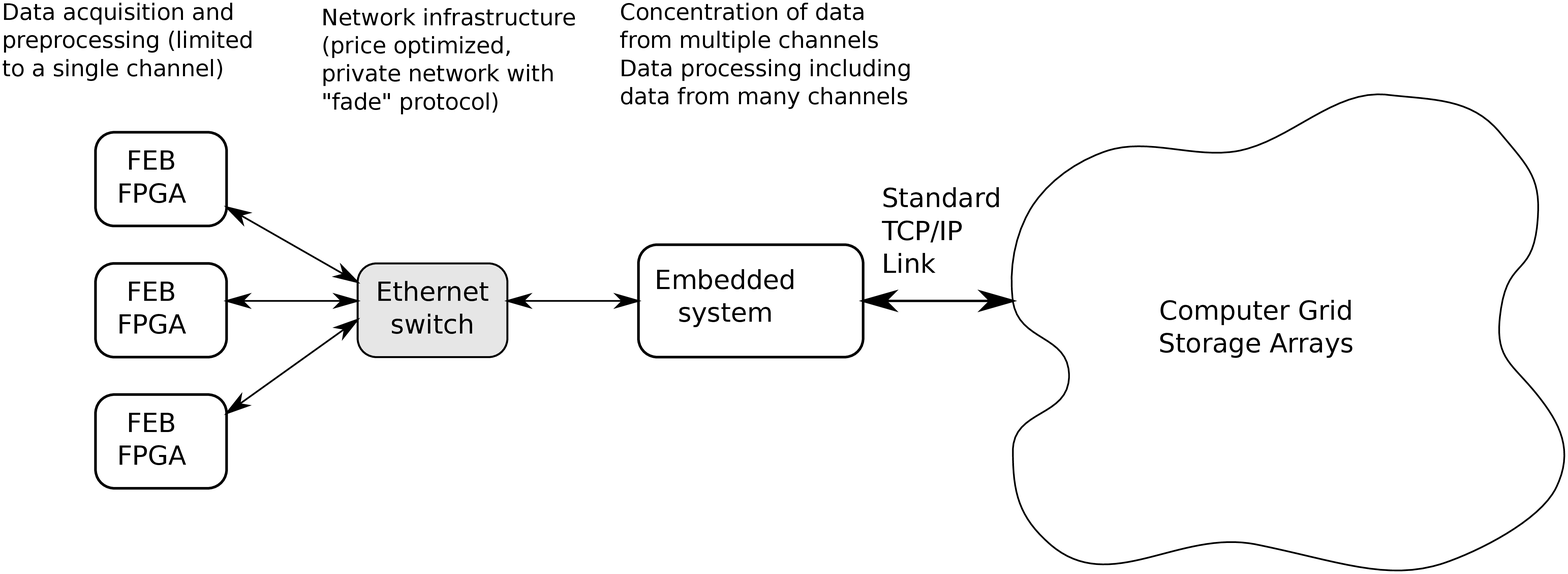}
   \end{tabular}
   \end{center}
   \caption[example]
   { \label{figure:ether-daq-1}
     Data acquisition system with multiple FEB connected via Ethernet switch
     to a single embedded system, transmitting
     data to the computer grid for further processing or storage.
   }
\end{figure}

The most popular standard used in local area networks (LAN) is the Ethernet, and therefore
we have focused on application of the Ethernet network to build a link between FEB boards
and the computer system. Of course this computer system may further distribute the concentrated
data to the whole data processing network, using the standard network interface as shown
in the Figure~\ref{figure:ether-daq-1}.

\section{Use of Ethernet interface with FPGA chips}
The Ethernet interface assures high throughput and reasonable latency, however it
does not assure reliable transfer of data.
In typical applications Ethernet is used as a network interface to 
provide low-level communication, and reliability
of the communication is assured by the higher layers of the network protocol.

The most popular protocol for reliable transmission of data through unreliable
network connection is TCP/IP, but for our application it is unacceptable.
The implementations of TCP/IP stack in the FPGA require a lot of 
resources~\cite{url-xlx-tcpip, url-xlx-lwip}, as it is necessary to provide CPU 
functionalities, and additionally a lot of memory to buffer
the data. 

The benefits of using TCP/IP, such as routability and possibility to transmit
the data via heterogeneous network (implemented in the IP layer, the $3^{rd}$ layer in the OSI model)
are useless for this specific application, where we need to transmit data in a single
network segment or in a few segments connected via Ethernet switch.

To find the optimal solution, we must quickly analyze consequences of requirement of reliable transmission.

\subsection{Reliable transmission}
\label{sec:reliable-transmission}
The Ethernet link is not reliable due to following reasons:
\begin{itemize}
 \item Data corruption in the physical link (the bit error rate (BER) may be up to
 $10^{-8}$ in 10Base-T~\cite{ieee-802-3-2008-p3-s1} and up to $10^{-10}$ in 100Base-T~\cite{ieee-802-3-2008-p3-s2}
 and 1000Base-T Ethernet~\cite{ieee-802-3-2008-p3-s3})
 \item Packet corruption due to collision (only in certain kinds of physical connections)
 \item Dropping of packets by the receiving system due to overload 
\end{itemize}
Infrequent random loss of a single packet (e.g. due to noise or collision) can be mitigated by using
Forward Error Correction~\cite{rfc3453} techniques, implemented on the protocol level. 
The simplest (and therefore implementable in a FPGA) method could be grouping of packets
in $N$-packet groups and sending an additional \WZcode{parity packet}, filled with data calculated as
exclusive-or of data contained in normal packets. If any single packet in the group is corrupted
(which can be detected using the packet's checksum), it may be reconstructed from other packets and the 
\WZcode{parity packet}\footnote{Similar solution was used to protect configuration data in the FLASH memories
against radiation induced corruption in the Linkbox Control
System for RPC subdetector in the LHC experiment~\cite{rlbcs-spie-2005}.}.
Such a solution may be worth of checking for a full-duplex connection
between single FEB and the network adapter in the computer system, as such configuration
minimizes packet collisions.  
Unfortunately it may be not effective in a typical situation where we want to concentrate data 
(with a few FEBs connected via network switch to the network adapter in the computer system),
as in this setup there is a higher risk
of dropping of more packets from a group.

In such situation, the only way to assure reliability is to use the acknowledge/retransmission mechanism
similar to the one used in the TCP/IP protocol.
The problem with such solution, however, is that it requires significant amount of memory to buffer the
transmitted, and not confirmed yet data.

To fully utilize the throughput of the network link, when waiting for the acknowledgment
of the particular packet, we should transmit next ones. However all unacknowledged packets
should be stored in the memory, because they may need to be retransmitted if no acknowledgment
is received.

If we denote the transmission speed as $\TransmSpeed$, and the maximum latency of acknowledgment
as $\MaxAckLatency$, then the required capacity of the memory buffer $\BufferCapacitance$ is given by
the simplified formula (the formula does not take into account the length of the packet):
\begin{equation}
\label{equ:transm1}
\BufferCapacitance =  \TransmSpeed~\MaxAckLatency
\end{equation}

Currently the typical amount of internal memory in inexpensive FPGA chips is below 100~KiB,
and if we consider the Ethernet protocol overhead, we can state that for transmission speed of 100~Mb/s
we need maximum acknowledge latency below c.a. 7~ms, and for transmission speed of 1~Gb/s below c.a. 700~µs.

The acknowledge latency measured for TCP/IP with direct connection (via switch only) 
is typically below that value. The measured mean ACK latencies were: 
\begin{itemize}
 \item 170~µs for communication with Intel Core 2 T5500/1.66~GHz based system
 \item 240~µs for communication with Pentium~4/2.8~GHz based system
 \item 520µs for Ralink RT3350/320~MHz based system.
\end{itemize}
 
However the latency increases above 1~ms, when routing is involved.

These facts show, that if we want to assure reliable transmission from FPGA with low amount
of internal memory, we can't use routing. The acknowledge must be generated by the receiving host
which is connected either directly, or at most via Ethernet switch.
However in this situation we also don't need an IP layer,
as for communication between devices connected via Ethernet switch, the MAC addressing is sufficient.

Summarizing the above analysis:
\begin{itemize}
 \item We can use simple, unroutable protocol using the MAC addressing.
 \item We should concentrate on minimizing of the acknowledge latency.
\end{itemize}

If the data received from the FPGA based FEBs should be transmitted further (see Figure~\ref{figure:ether-daq-1}),
we can use the standard network solutions for that next stage of transmission (of course the required link throughput
for this connection may be higher, than for links from FEBs).
As the typical Linux based embedded computer system is equipped with RAM memory with capacity of
at least 64~MiB and often above 1~GiB, it is not a problem to buffer much higher amount of data 
than in the FPGA. Therefore, according to the formula (\ref{equ:transm1}), 
further transmission may be performed with routable protocols
with higher acknowledge latency, like TCP/IP.

\subsection{Avoiding of network congestion}

As it was stated above, one reason why Ethernet does not assure reliable transmission, is the danger
of packet loss due to collisions or due to dropping of packets by overloaded switch or receiving system.
Systems, which use packet acknowledge mechanisms are prone to the network congestion problem,
when quick resending of not acknowledged packets leads to increase of network load and receiver load, which
in turn further increases the risk of packet loss.

To avoid the network congestion, the system should be equipped with means to monitor the ratio of the lost
and retransmitted packets and to adjust the rate of sending of packets, so that the amount of
dropped packets is reasonable.
As the implementation described in this paper
is only a ``proof of the concept'' solution, we have proposed very simple mechanism
(described in subsection \ref{sec:nca}),
where the delay  between packets is adjusted depending on the ratio of lost packets.
Further research is needed
to find the optimal method to avoid switch or receiver overload. The problem maybe especially important
in triggered data acquisition systems, where all FEBs may start to transmit data simultaneously, after
the trigger is received and the network load is fluctuating.

\subsection{Additional assumptions simplifying the design}

The system is supposed to work over the private, physically protected network.
Therefore we don't need all the features, which are introduced in protocols like TCP/IP to assure
secure communication (e.g. we can use simple sequential numbering of packets).
Additionally at this, initial, state of development there was no need to officially allocate
the Ethernet protocol number, and an arbitrarily chosen number \WZhex{0xfade} was used.

\section{Implementation of the system}
Next sections describe the implementation of all parts of the proposed system, including the proposed 
communication protocol, the FPGA IP core, and the Linux kernel driver for the embedded system.

\subsection{Proposed communication protocol}

Design of the communication protocol heavily depends on the details of FPGA implementation 
(see section~\ref{sec:fpga-implementation}) 
and of software implementation (see section~\ref{sec:software-implementation}).
Therefore describing the protocol we will often mention details which are fully explained in the next 
sections.

The communication protocol is kept as simple as possible. Hence we use only four
kinds of packets (see Table~\ref{tab:packets}.)
\begin{table}[tp]
   \caption[example]
   { \label{tab:packets}
     Structure of the packets used by the transmission protocol.
     (SRC and TGT - MAC addresses of the transmitter and of the receiver)
   }
   {\small
\begin{tabular}{|p{2cm}|c|c|}
\hline
packet & direction & structure \\
\hline\hline
 & & \\
START packet & to FEB  &
 \begin{tabular}{|c|c|c|c|c|}
   \hline
   SRC & TGT & 0xfade & 0x0001 & padding to 64 bytes \\ 
   \hline
 \end{tabular}
 \\ & & \\
STOP packet & to FEB  &
 \begin{tabular}{|c|c|c|c|c|}
   \hline
   SRC & TGT & 0xfade & 0x0005 & padding to 64 bytes \\ 
   \hline
 \end{tabular}
 \\ & & \\
DATA packet & from FEB &
 \begin{tabular}{|c|c|c|c|p{2cm}|c|p{2cm}|}
   \hline
   SRC & TGT & 0xfade & 0xa5a5 & set number \& packet number & delay & 1024 bytes of data \\ 
   \hline
 \end{tabular}
\\ & & \\
ACK packet & to FEB  &
 \begin{tabular}{|c|c|c|c|p{2cm}|p{2cm}|}
   \hline
   SRC & TGT & 0xfade & 0x0003 & set number \& packet number & padding to 64 bytes \\ 
   \hline
 \end{tabular}
 \\ & & \\
\hline
\end{tabular}
}

\end{table}

To start and stop transmission of data from the FEB, the receiving computer
system sends appropriately the START or STOP packet.

After the transmission is started, the FEB starts to send the stream of data.
Data bytes are encapsulated in DATA packets containing 1024 bytes of data and
an additional information (described further).

The length of packet was chosen so, that it is shorter than the Ethernet
Maximum Transmission Unit (MTU), equal to 1518 bytes\cite{ieee-802-3-2008-p3-s1}, 
and is equal to power of two, which allows efficient storage of packet contents
in the memory buffer.

The data stream is logically divided into \WZcode{data sets}. 
One \WZcode{data set} is a group
of consecutive packets, carrying the data, which fit into the memory used to buffer the
transmitted and not confirmed yet packets. 
Due to the technical characteristics of FPGA platforms available for tests
(see section~\ref{sec:tests-and-results}), the length of the \WZcode{data set} was chosen
as $\NOfPktBuffers=32$. The number of packets in a set is also a power of two,
which simplifies addressing of the memory.

The transmitted packets are labeled with the sequential number, created by the 
concatenation of the 16-bit set number and 5-bit packet number\footnote{
In fact current FPGA sources reserve place for 6-bit packet number, as some of tested
FPGA chips allow to use longer sets. Additionally for debugging purposes it is possible to include
the 10-bit retry number (currently not used) in each packet.} 

After reception of the DATA packet, the receiving system confirms the successful reception,
sending the ACK packet, labeled with the same set number and packet number as the acknowledged
packet.

The proposed protocol may be extended with additional command packets, sent from the embedded 
system to the FPGA, providing more advanced control of the FEB's operation.
Such extension, however, implies necessity to introduce the acknowledge/retransmit mechanism
for the command packets as well.
(Currently implemented commands START and STOP do not require acknowledgement, as their correct
reception is confirmed by sending or not sending of data by the FEB).

\subsection{FPGA implementation}
\label{sec:fpga-implementation}
\begin{figure}[tp]
   \begin{center}
   \begin{tabular}{c}
   \includegraphics[width=0.8\linewidth]{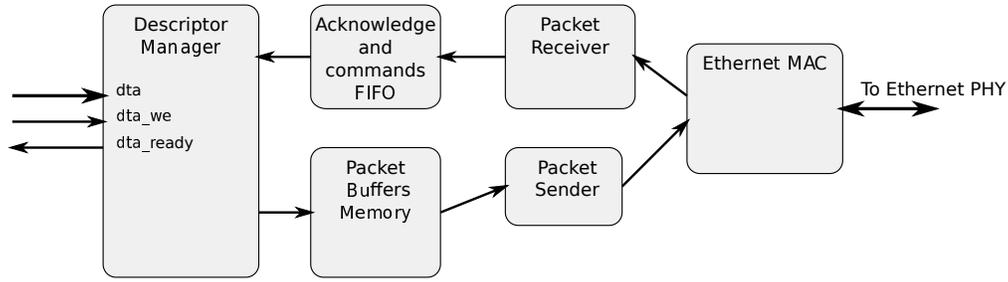}
   \end{tabular}
   \end{center}
   \caption[example]
   { \label{figure:fpga-ip-core}
     Structure of the FPGA IP core implementing the hardware part of the system.
   }
\end{figure}

Structure of the FPGA IP core is shown in the figure~\ref{figure:fpga-ip-core}

The input of the IP core behaves like typical FIFO input. The \WZcode{dta\_ready} signal informs if the core is
ready to accept the new data. The \WZcode{dta} signal is a 32-bit wide data bus. The \WZcode{dta\_we} signal is the 
data write strobe.

Main part of the IP core is the subsystem which manages transmission and retransmission of packets
(the \WZcode{Descriptor Manager} - further denoted as DM),
and stores the packets (the \WZcode{Packet Buffers Memory} - further denoted as PBM).

\begin{figure}[tp]
   \begin{center}
   \begin{tabular}{c}
   \includegraphics[width=0.8\linewidth]{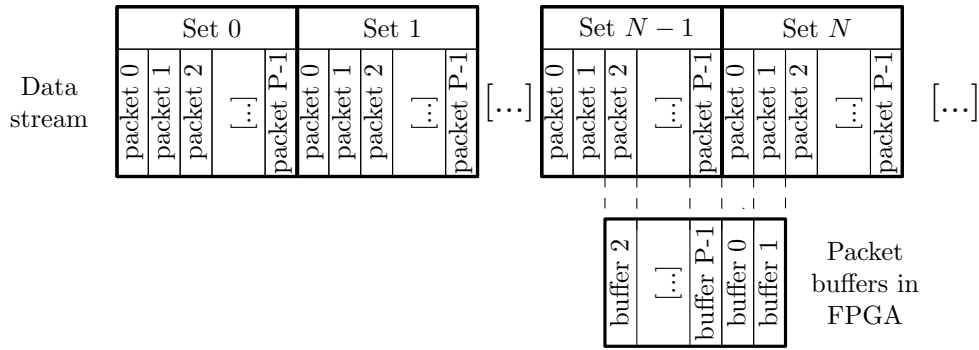}
   \end{tabular}
   \end{center}
   \caption[example]
   { \label{figure:packets-and-buffers}
     Relation between packets and packet buffers. Each packet set contains $P=\NOfPktBuffers$ packets numbered 
     from $0$ to $P-1$.
   }
\end{figure}
The PBM, used to buffer the transmitted data,
is divided into $\NOfPktBuffers=32$ packet buffers, each 1024 bytes long.
The PBM works as a circular buffer. The $i^{th}$ packet of any set is always stored in the $i^{th}$ 
packet buffer. The PBM always stores packets belonging to one \WZcode{data set} or to two
 neighboring \WZcode{data sets} 
(see Figure~\ref{figure:packets-and-buffers}).

Information about the state of each packet buffer is stored in the corresponding 
\WZcode{packet descriptor} record stored in the DM.

Each descriptor stores:
\begin{itemize}
 \item {three bit flags, describing the state of the packet:
   \begin{itemize}
     \item {\em valid (V)} - set, when the packet is filled with the data,
     \item {\em sent (S)} - set, when the packet has been transmitted at least once,
     \item {\em confirmed (C)} - set, when the reception of the packet has been 
           confirmed by the receiver.
   \end{itemize}
  Initially all those flags are set to 0. }
 \item The \WZcode{set number} - informing to which \WZcode{data set} belong the data stored in the 
       particular packet buffer
\end{itemize}

The table of \WZcode{packet descriptors} is handled as a circular buffer by the DM. 
The \WZcode{head} pointer points to the packet buffer which is currently
being filled with the data.
The \WZcode{tail} pointer points to the last filled packet buffer,
which has not been confirmed yet by the receiver, or is equal to the \WZcode{head} pointer, if 
there are no packet buffers ready for transmission.
The third \WZcode{retr} pointer points to the packet buffer,
which should be transmitted or retransmitted when the transceiver
is ready, and when the required delay between packets 
(see section~\ref{sec:nca}) has elapsed.

The main state machine in the DM handles three different tasks (in order of decreasing priority):
\begin{itemize}
 \item reception of acknowledge packets, and handling of \WZcode{tail} pointer
 \item detection of completely filled packets and handling of \WZcode{head} pointer
 \item transmission and retransmission of packets and handling of \WZcode{retr} pointer
\end{itemize}

Whenever the acknowledge packet is received, the state machine checks
if the \WZcode{set number} in the acknowledge packet agrees with the \WZcode{set number}
in the corresponding packet descriptor. If the \WZcode{set numbers} do not agree, such
acknowledge packet is discarded. If the \WZcode{set numbers} agree, the C flag is set to 1.
Additionally, if the acknowledged packet buffer is the one pointed by the \WZcode{tail} pointer,
we start moving the \WZcode{tail} pointer,
until it points to the first sent but not confirmed packet (S=1 and C=0)
or is equal to the \WZcode{head} pointer (meaning that all data are transferred).

If the new data are delivered to the input, they are written to the packet buffer
pointed by the \WZcode{head} pointer.
When this packet buffer is completely filled, it is marked as ready for transmission
(by setting V=1), and the signal \WZcode{dta\_ready} is cleared, signaling, 
that the DM is not ready
for new data.

The state machine detects this state, and tries to move the \WZcode{head}
pointer to the next position. If the next position is still pointed by the \WZcode{tail} pointer,
it means, that there is no place for new data, and we must wait until the data are transmitted
and confirmed. 
If the next position is free, the \WZcode{head} pointer is moved
and the signal \WZcode{dta\_ready} is set,
allowing to feed the DM with the next data. 

Whenever the \WZcode{head} pointer is moved, the flags V, S and C in the descriptor pointed by the
new \WZcode{head} value are cleared, and the \WZcode{set number}
in this descriptor is set to the number of currently
transmitted \WZcode{data set}.

Finally the state machine checks, if the \WZcode{retr} pointer points to the packet buffer which is valid,
but not confirmed (V=1 and C=0). If yes, it orders transmission or retransmission of the corresponding packet, sets
the S flag in its packet descriptor, and moves the \WZcode{retr} pointer to the next packet descriptor
ready for transmission or retransmission (V=1 and C=0). If \WZcode{retr} pointer reaches the \WZcode{head} position
it is wrapped to the \WZcode{tail} position. If the buffer is empty
(\WZcode{head} = \WZcode{tail}) no transmission 
is attempted\footnote{This simplified description does not cover special situation, when just confirmed packet
is the one pointed by the \WZcode{retr} pointer. In this situation the \WZcode{retr} 
pointer must be also modified.}.

Blocks \WZcode{Packet sender} and \WZcode{Packet receiver} are designed to autonomously service transmission
and reception of the data. 
They are introduced to parallelize reception of the data from the system input 
(associated with packet buffers management), and Ethernet transmission.
Additionally those blocks (together with the \WZcode{Acknowledge and Commands FIFO}) 
allow to separate clock domains between the Ethernet related part of design and the rest of the system.

To control the Ethernet PHY our system uses the open source Ethernet MAC~\cite{opencores-10-100-1000-mac}, 
however it can be also easily adapted to work with another MAC or to directly communicate with the Ethernet PHY.

\subsubsection{Implementation of network congestion avoidance in FPGA}
\label{sec:nca}
To avoid network congestion, caused by too high frequency of packet transmission, resulting
in dropping of packets by the switch or by the receiving system, we have introduced adjustable
delay between transmitted packets. 

Because selection of the proper delay in advance is difficult or even impossible,
we have introduced simple mechanism to adapt this delay. 

Transmission starts with 
the delay set to the high value, equal to 200µs, which should minimize number of lost packets, at cost 
of suboptimal utilization of the link throughput.
During transmission the FPGA measures the ratio between the number of all transmitted 
packets and the number of retransmitted packets and adjusts the delay accordingly.

Whenever a packet is sent or resent, we increase the counter of sent packets ($\NOfSentPackets$).
If the transmitted packet is resent (S=1 in descriptor flags) additionally we increase the counter
of resent packets ($\NOfResentPackets$).

After the pre-defined number of packets is sent ($\NOfSentPackets = \NOfCountedPackets$),
we check the ratio of the resent packets $\RatioResent=\NOfResentPackets/\NOfSentPackets$, and compare it with
two pre-defined thresholds: $\NCAHighThreshold$ and $\NCALowThreshold$. 

If the ratio of resent packets is higher then $\NCAHighThreshold$, the delay between packets
is multiplied by the factor $\NCAIncreaseDelay > 1$. If the ratio of resent packets lower then
$\NCALowThreshold$, the delay between packets is multiplied by the factor $\NCADecreaseDelay<1$.
Afterwards both counters ($\NOfResentPackets$ and $\NOfSentPackets$) are cleared.

Tests were performed for two different sets of values of the parameters:
\begin{itemize}
 \item $\NOfCountedPackets=3000$, $\NCAHighThreshold=1/16$, $\NCALowThreshold=1/64$, $\NCAIncreaseDelay=1.25$,
$\NCADecreaseDelay=0.9375$.
 \item $\NOfCountedPackets=10000$, $\NCAHighThreshold=1/8$, $\NCALowThreshold=1/32$, $\NCAIncreaseDelay=1.25$,
$\NCADecreaseDelay=0.75$
\end{itemize}
In tests (see section~\ref{sec:tests-and-results}) both sets of settings provided reliable operation.

\subsection{Implementation of the Linux driver}
\label{sec:software-implementation}
In the embedded system the communication with FEBs is handled by a device driver, 
working in the Linux kernel space, which may be controlled by the user space application.
The user space application may receive acquired data in an efficient
way, using the memory mapped circular buffer.
Such approach simplifies implementation  of algorithms of data preprocessing and further distribution,
as the user space code is easier to develop and debug than the kernel code.
Use of a memory mapped buffer, instead of traditional socket interface, to 
access received data from the user space, allows to decrease overhead needed to handle data.

The driver may service multiple FPGA based FEBs, and multiple network interface cards (with possibility
to have a few FEBs connected via a switch to one network card). The maximum number of FEBs
is defined by the module parameter \WZcode{max\_slaves}.
For each FEB a separate character device 
(\WZcode{/dev/l3\_fpga0}, \WZcode{/dev/l3\_fpga1} and so on)
is created with separate circular buffer. The data in each buffer may be accessed directly
using \WZcode{mmap} technique, but the pointers position in the particular 
buffer must be accessed using the \WZcode{ioctl} function, to assure proper synchronization.
Such approach assures both fast and secure access to the received data.

The main component of the device driver is the protocol handler, installed via \WZcode{dev\_add\_pack}
function \cite[chap. 13]{underst-linux-network-internals},
which is called whenever the Ethernet frame with \WZhex{0xfade} type is received.
The protocol handler first checks if the received packet has been transmitted by the
registered and started FEB (and if no, it sends the STOP command to the source of the packet).
Then it checks if the received packet has reasonable \WZcode{set number}. If the \WZcode{set number} corresponds
to the already confirmed packet, the acknowledgment is sent immediately to handle 
situation when the previous acknowledge packet got lost. If the \WZcode{set number} corresponds
to the packet, which has not been received yet, the handler tries to copy received data to the circular
buffer, and if it succeeds, marks the packet as confirmed and sends the acknowledgment packet.

Due to possible loss of packets, it is not warranted, that the data arrive in sequence. Therefore 
the \WZcode{head} pointer is moved only to the end of the continuous area filled with the received data.

To speed up reception of the data, some optimizations are undertaken. The length of the circular
buffer is a multiple of the length of single \WZcode{data set} ($n \cdot \NOfPktBuffers \cdot 1024$ bytes),
and therefore the data associated with particular \WZcode{data set}
always occupy a continuous area in the circular buffer. 
As we always expect packets only from two consecutive \WZcode{data sets}
(see Figure~\ref{figure:packets-and-buffers}), it is enough to maintain two pointers, pointing
to the beginning of data associated with those \WZcode{data sets} in the circular buffer.

Described optimizations allow to minimize the time needed to copy the data and to acknowledge
the packet, which in turn allows to achieve faster transmission, according to formula \ref{equ:transm1}.

\subsubsection{Communication with the user space application}
The user space application may connect to one or more FEB devices, opening one or more created 
character devices (\WZcode{/dev/l3\_fpga0}, \WZcode{/dev/l3\_fpga1} and so on), and mapping its
circular buffer into the application's address space.

The application may connect to the particular FEB, and start transmission, using the special
ioctl call: \WZcode{L3\_V1\_IOC\_STARTMAC}, passing address of the  structure 
describing the desired FEB device (containing
its MAC address, and the name of the network interface - e.g. ``eth0'').

After the FEB device is connected, the system will start to receive the data, writing them to the circular buffer.
The application may read the \WZcode{head} and \WZcode{tail} pointers for the particular FEB device with
another ioctl call \WZcode{L3\_V1\_IOC\_READPTRS}, which additionally returns the number of available bytes
of data. Then the application may process the data located between the \WZcode{tail}
and \WZcode{head} position. After data are processed, the application may confirm their reception,
calling the \WZcode{L3\_V1\_IOC\_WRITEPTRS} ioctl with number of processed bytes.

Data located between the  \WZcode{tail}
and \WZcode{head} positions are warranted to be unchanged, so they maybe safely read and processed
by the application, while the \WZcode{ioctl} function takes care of proper synchronization
of access to the pointers between the application and the protocol handler.

To optimize use of the CPU, the application may sleep, waiting for the data.
To allow servicing of multiple FEB devices, the sleep functionality has been
implemented in the \WZcode{poll} function.
The number of available received bytes, needed to wake up the application may be defined with ioctl
\WZcode(L3\_V1\_IOC\_SETWAKEUP) call.

Additional ioctl commands allow to stop the transmission from particular
FEB device (\WZcode{L3\_V1\_IOC\_STOPMAC}), and to read the total length of the particular circular buffer 
(\WZcode{L3\_V1\_IOC\_GETBUFLEN}).

\subsubsection{Code portability}
The Linux device driver was prepared for Linux kernels 3.3.x , but it compiles also with newer kernels. 
Particularly it has been successfully compiled and used with Linux kernels 3.4.4 (in Knoppix 7.0.3~\cite{url-knoppix-703}),
3.5.2 and 3.5.3 (with Debian/testing). The code has been implemented in a multiprocessor-safe way.

\section{Tests, results and discussion}
\label{sec:tests-and-results}
The system was tested using the Dell Vostro 3750 (Intel Core i7-2630QM CPU, 2.0~GHz clock)
computer running the Debian/testing Linux OS (simulating the embedded system).
Use of computer with 4-core CPU (with hyperthreading capable cores) allowed to confirm, that 
the code works reliably in multiprocessor environment.
The FPGA based FEBs were simulated with three evaluation boards:
\begin{itemize}
\item SP601 evaluation board\cite{url-xlx-sp601} equipped with 10M/100M/1G Ethernet PHY
\item Atlys board\cite{url-atlys} equipped with 10M/100M/1G Ethernet PHY
\item Spartan-3E Starter Kit\cite{url-xlx-spart-3e-sk} (further denoted as S3ESK) equipped with 10M/100M Ethernet PHY
\end{itemize}
\begin{figure}[t]
   \begin{center}
   \begin{tabular}{c}
   \includegraphics[width=0.8\linewidth]{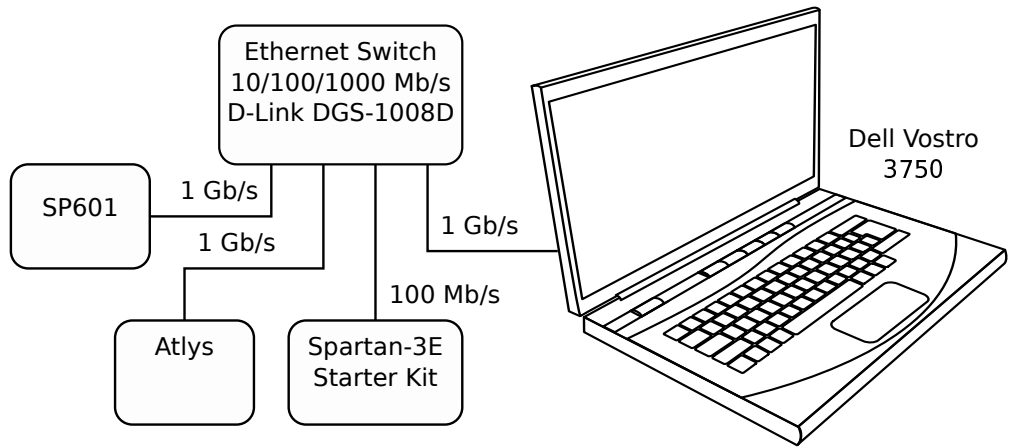}
   \end{tabular}
   \end{center}
   \caption[example]
   { \label{figure:test-setup}
     Test setup.
   }
\end{figure}
\subsection{Results of FPGA compilation}
The IP core was successfully compiled for all FPGA platforms used for tests.
The resources consumption for different platforms is shown in the Table~\ref{table:compilation-results}.
As can be seen, in all tested FPGAs, synthesis of our IP core leaves 
significant amount of resources for another, user defined functionalities.

Probably resources consumption may be further decreased by replacement of the OpenCores
MAC core \cite{opencores-10-100-1000-mac} with a simplified core, communicating directly with the Ethernet PHY.
Unfortunately such implementation is yet not mature enough to be published.

\begin{table}[t]
\caption{
\label{table:compilation-results}
Results of compilation of the IP core for different platforms}
\begin{tabular}{|c|c|c|c|}
\hline
Board & FPGA chip & Slice usage & RAM usage \\
\hline\hline
SP601 & xc6slx16 &  487 out of 2278 (21\%) &  RAMB16BWER: 20 out of 32 (62\%) \\
Atlys & xc6slx45 & 507 out of 6822 (7\%) & RAMB16BWER: 20 out of 116 (17\%) \\
S3ESK & xc3s500e & 1510 out of 4656 (32\%) & RAMB16: 20 out of 20 (100\%) \\
\hline
\end{tabular}  
\end{table}

\subsection{Results of transmission tests}
The tests were performed using a simple application, which received data sent by the emulated FEBs
and immediately confirmed their reception, freeing the buffer. 

The tests covered:
\begin{itemize}
 \item measurement of the throughput
 \item capture of packets sent and received by the computer, and analysis of the acknowledge latency and ``delay'' value
       reported by the emulated FEBs in the data packets (using the wireshark\cite{url-wireshark} tool)
\end{itemize}

The tests were performed with the 32~KiB of internal RAM in each FPGA used for packet buffers (32 packets in a single set).
All tested boards were connected to the 10/100/1000~Mb/s Ethernet switch D-Link DGS-1008D~\cite{url-dlink-dgs-1008d}
(see Figure~\ref{figure:test-setup}.),
and each board was assigned a unique MAC address. Results presented in the Table~\ref{table:tests-transmission-speed} were obtained for parameters
of the network congestion avoidance (NCA) algorithm set to $\NOfCountedPackets=3000$, $\NCAHighThreshold=1/16$, $\NCALowThreshold=1/64$, $\NCAIncreaseDelay=1.25$,
$\NCADecreaseDelay=0.9375$.
The transmission speed was measured during 5 minutes to minimize influence of initial adjustment of the inter-packet delay.
For each combination of active boards 12 to 15 measurements were performed, and both average and standard deviation of
transmission rate was calculated.
\newcounter{rowNoTests}
\begin{table}[tp]
\caption{
\label{table:tests-transmission-speed}
Results of measurement of efficient transmission speed}
{\small
\begin{tabular}{|>{\refstepcounter{rowNoTests}}c|c|c|c|c|}
\hline
\multirow{2}{*}{Active boards} & \multicolumn{4}{|c|}{Measured efficient transmission speed [Mb/s]} \\
\cline{2-5}
 & SP601 (1~Gb/s) & Atlys (1~Gb/s) & S3ESK (100~Mb/s) & Total \\
\hline\hline
\setcounter{rowNoTests}{1}
All boards active & 261.42 ($\sigma$=14.58)&  284.44 ($\sigma$=13.44)& 94.86 ($\sigma$=0.36) & 640.72 ($\sigma$=26.17)\\
\label{row:atl-sp601}SP601 and Atlys active & 306.09 ($\sigma$=12.03)&  331.44 ($\sigma$=18.70)& -- & 637.53 ($\sigma$=29.97)\\
SP601 and SK3E & 481.52 ($\sigma$=18.92) &  -- & 94.98 ($\sigma$=0.016) & 576.51 ($\sigma$=18.93)\\
Atlys and SK3E & -- & 515.27 ($\sigma$=19.21) &  94.99 ($\sigma$=0.017) & 610.25 ($\sigma$=19.21)\\
\label{row:sp601}SP601 alone & 543.29 ($\sigma$=13.40) & -- & -- & 543.29 ($\sigma$=13.40)\\
\label{row:atl}Atlys alone & -- & 569.95 ($\sigma$=14.12) & -- & 569.95 ($\sigma$=14.12)\\
SK3ESK alone & -- & -- & 95.02 ($\sigma$=18.99) & 95.02 ($\sigma$=18.99)\\
\hline
\end{tabular}  
}
\end{table}
Presented results show, that the proposed system reasonably utilizes (over 50\%) 
the bandwidth of the 1~Gb/s Ethernet link available in the receiving system. Probably the achieved throughput
could be limited by the speed, at which the application in the receiving system was able to verify the data.

Achieved total throughput is higher for two FEBs with 1~Gb/s interface 
(Table~\ref{table:tests-transmission-speed}, row~\ref{row:atl-sp601}), than for one such board 
(Table~\ref{table:tests-transmission-speed}, row~\ref{row:sp601} and row~\ref{row:atl}).
This probably may be explained by the fact, that the total amount of memory, used to buffer data at
the transmitting side is higher in case of two FEBs.
The SP601 board provided slightly smaller transmission rate than the Atlys board, even though both boards
are equipped with 1Gb/s Ethernet interface, and this result is not explained yet.
The SK3E board, equipped with 100~Mb/s Ethernet interface was able to transmit the data with practically maximum speed
in all configurations.

The second set of settings ($\NOfCountedPackets=10000$, $\NCAHighThreshold=1/8$,
$\NCALowThreshold=1/32$, $\NCAIncreaseDelay=1.25$,
$\NCADecreaseDelay=0.75$) of the NCA algorithm was also tested, and appeared to provide reliable transmission of data.

The mean acknowledge latency, measured with the wireshark tool, was equal to 3~µs, and was significantly lower than the latencies
measured in the same computer system for the TCP/IP protocol (see section~\ref{sec:reliable-transmission}).

Additional tests, with delay introduced in the user space application, proved that the FPGA IP core is able 
to successfully adapt delay between packets to the processing speed of the receiving system.

The tests have shown that presented system is able to reliably transfer the data
from multiple FPGA based FEBs connected via an Ethernet switch to the network interface
card in the receiving computer system.

The applied, very simple, network congestion avoidance mechanism appeared to
work reliably in case of single, two or three FEBs sending the continuous stream of data,
even if the data rates produced by those FEBs differ significantly.

Probably further research may be needed to investigate 
reliability of our NCA algorithm in more difficult conditions - e.g. in the situation typical
for triggered data acquisition systems, where the data rate is fluctuating, reaching the peak value
right after the trigger.

\section{Availability of the code}
The first released version of the code, implementing the described system,
has been announced \cite{l3fade-comp.arch.fpga}
and published \cite{l3fade-alt.sources} in the Usenet newsgroups.
The newest version, including files needed to implement it on boards SP601\cite{url-xlx-sp601}, 
Atlys\cite{url-atlys} and Spartan-3E Starter Kit\cite{url-xlx-spart-3e-sk}
is available on the dedicated website \cite{url-l3-fade}.

The licensing information is included in the archive, but generally the whole system is freely
available as the open source code, partially under the GPL license, partially under the BSD license,
and partially as public domain.

\section{Conclusions}
The presented system, consisting of the dedicated FPGA IP core, the simple network protocol and 
the specialized Linux device driver,
allows to reliably transmit data from FPGA based Front End Boards (FEBs), to the embedded system
via an Ethernet link.

Due to simplicity of the proposed protocol, which leads to simple implementation of the FPGA IP core, 
and due to minimization of the packet acknowledge latency in the device driver, the system allows
to obtain fast, reliable transmission even for small and inexpensive FPGA chips, without necessity to
connect them to external RAM.

Data received by the embedded system are placed in the circular buffer, directly available
(with memory mapping, assuring minimal data access overhead) 
for the user space application, which may quickly process and further distribute them.
The processing speed may be further increased by use of a multiprocessor embedded system.

In tests the described system allowed to reliably transfer data from 3 FPGA based FEBs,
with total throughput up to c.a. 640~Mb/s, 
via 1Gb/s Ethernet link, using only 32~KiB of internal FPGA memory for data buffering.

The proposed system may be used to concentrate data from FPGA based FEBs
to the data processing network, using the standard, inexpensive components,
like Ethernet cables and switches, and embedded computer systems equipped with multiple network adapters.

\bibliographystyle{model1-num-names}   %
\bibliography{report}   %

\vspace{1cm}
~\\
\begin{tabular}{p{0.15\linewidth}p{0.8\linewidth}}
  \begin{minipage}[b]{\linewidth}
    \includegraphics[width=\linewidth]{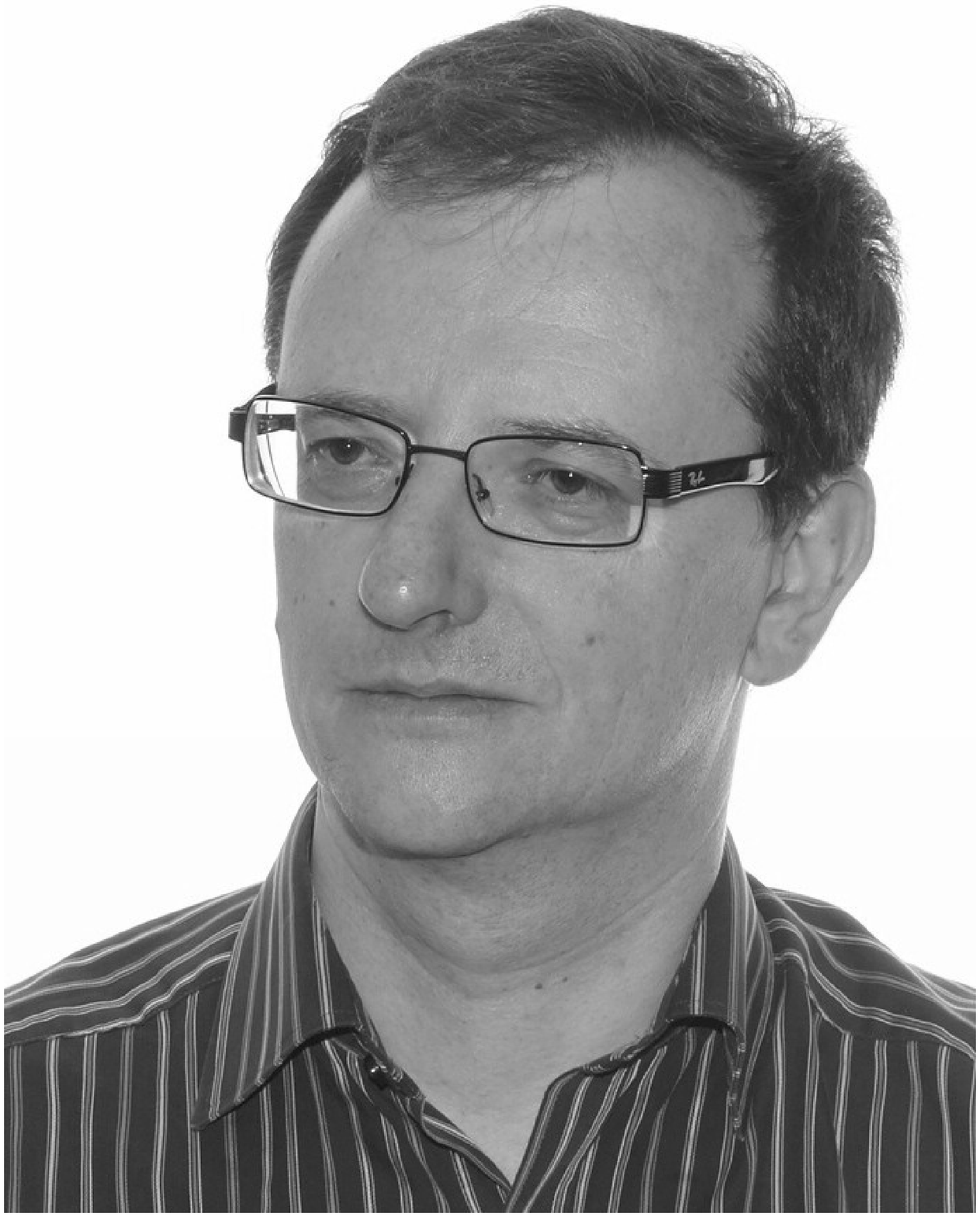} 
  \end{minipage}
&
  \begin{minipage}[b]{\linewidth}
     Wojciech M. Zabolotny was born in Sucha Beskidzka,
Poland in 1966. He received the MSc (1989)
and the Ph.D. (1999) in Electronics from the
Warsaw University of Technology in Poland, both with honors.
Since 1990 he was a research assistant and since 1999 he
is an Assistant Professor at the 
Warsaw University of Technology.
His research interests are the distributed data acquisition
systems (biomedical and for high energy physics),
the embedded systems and programmable logic.
He was involved in development of electronic systems for
CERN (since 2001), for DESY in Hamburg (2002-2009) 
and for JET in Culham (since 2010).

  \end{minipage}
\\
\end{tabular}
\end{document}